\documentclass[10pt]{article}
\usepackage{amsmath,amssymb,dsfont,graphicx}
\newcommand{\beq}{\begin{equation}}
\newcommand{\eeq}{\end{equation}}
\newcommand{\ba}{\begin{array}}
\newcommand{\ea}{\end{array}}
\newcommand{\bea}{\begin{eqnarray}}
\newcommand{\eea}{\end{eqnarray}}
\newcommand{\bean}{\begin{eqnarray*}}
\newcommand{\eean}{\end{eqnarray*}}

\newtheorem{proof}{Proof.}

\newcounter{appendix}
\newcommand{\half}{\frac{1}{2}}

\def\mat2#1#2#3#4{{\left(\begin{array}{cc}#1 & #2\\ #3 & #4
      \end{array}\right)}}

\def\mats2#1#2#3#4{{\left(\begin{array}{cc}#1 & #2\vspace{2truemm} \\ #3 & #4
\end{array}\right)}}
\begin{document}
\begin{titlepage}
\begin{center}
{\Huge Discrete Reductive Perturbation Technique}
\end{center}
\vspace{0.8truecm}
\begin{center}
{\Large
Decio Levi$^\diamondsuit$ and Matteo Petrera$^\sharp$}
\vskip0.8truecm

$^\diamondsuit$  Dipartimento di Ingegneria Elettronica, \\
Universit\`a degli Studi Roma Tre and Sezione INFN, Roma Tre,\\
Via della Vasca Navale 84, 00146 Roma, Italy.\\
E--mail: levi@fis.uniroma3.it\\
\vspace{0.8truecm}
$^\sharp$ Dipartimento di Fisica E. Amaldi \\
Universit\`a degli Studi di Roma Tre and Sezione INFN, Roma Tre,\\
Via della Vasca Navale 84, 00146 Roma, Italy.\\
E--mail: petrera@fis.uniroma3.it\\

\end{center}
\vspace{0.2truecm}
\vspace{0.2truecm}
\abstract{\noindent

We expand a partial difference equation (P$\Delta$E) on multiple
 lattices and obtain the P$\Delta$E which governs its far field behaviour. 
The perturbative--reductive approach is here performed on well known nonlinear P$\Delta$Es, 
both integrable and non integrable. We study the cases of 
the lattice modified Korteweg--de Vries (mKdV) equation, 
the Hietarinta equation, the lattice Volterra--Kac--Van Moerbeke (VKVM) equation and
a non integrable lattice KdV equation. Such reductions allow us to obtain many new P$\Delta$Es of the 
nonlinear Schr\"odinger (NLS) type.

}\vskip 1truecm \noindent
%Keywords:  \\
%PACS: 

\tableofcontents
\end{titlepage}

%%%%%%%%%%%%%%%%%%%%%%%%%%%%%%%%%%%%%%%%%%%%%%%%%%%%%%%%%%%%%%%%%%%%%%%%%%%%%%%%%%%%%%%%%%%%%%%%%%%%%%%%
%%%%%%%%%%%%%%%%%%%%%%%%%%%%%%%%%%%%%%%%%%%%%%%%%%%%%%%%%%%%%%%%%%%%%%%%%%%%%%%%%%%%%%%%%%%%%%%%%%%%%%%%
\section{Introduction}
Problems involving the evolution of nonlinear phenomena, both continuous and 
discrete, have become of increasing interest in various branches of science and 
engineering.
Nonlinear waves, without dissipation and dispersion give rise in a finite time to a 
discontinuity. A typical example of nonlinear wave is the shock wave produced by a 
supersonic object. Dissipation and dispersion play an important role in balancing the 
steepening due to nonlinearity, so that when these effect are present, a steep but 
smooth solitary wave may be formed and then propagates for all times. The solitary 
wave phenomenon has actually been observed for many years in the form of a surface 
wave in shallow water. A model equation of nonlinear dispersive phenomena may, in general, be very 
complicate. The soliton may appear only in the asymptotics, after a long 
transient period. Thus to be able to put in evidence the solitons, Taniuti and 
collaborators \cite{t1,t2} introduced an asymptotic method which makes it possible to reduce 
general nonlinear evolution equations to some more tractable nonlinear equations. 
This method go under the denomination of {\it{reductive perturbation technique}}. Under 
the assumption that the amplitude of the waves are small, one is able to reduce the 
starting hyperbolic system to a few simple equations, such as the Burgers equation, the 
Korteweg--de Vries equation, the nonlinear Schr\"odinger equation and few others. 

In the reductive perturbation method, the space and time coordinates are stretched in 
terms of a small expansion parameter and we introduce the concept of {\it{far field}},
as the field governing the asymptotic behaviour of the reduced equation.
To give a simple idea of the reasoning underlining this concept, let us consider, as an example, the 
familiar wave equation in two variables:
\beq \label{i1}
\phi_{,tt}-\phi_{,xx} = 0.
\eeq
The general solution of equation (\ref{i1})
can be expressed as the superposition of waves moving to the right 
and to the left. In general these two waves are excited simultaneously by an arbitrary 
initial condition. However, if the initial condition is localized, after a certain finite time
the disturbance separates in a progressive wave propagating to the right and one to the 
left, and they are solutions to a first order equation, i.e. an equation of one fewer degree of freedom:
\beq \nonumber
\phi_{,x} \pm \phi_{,t} = 0.
\eeq
We call the solutions of the first order equation the {\it{far field solutions}} of the original wave 
equation. The concept of far field came from the idea of finding properties of a given
evolution equation which do not depend in a sensitive manner on the details of the initial 
conditions, but correspond to a wide class of initial conditions. 

As an example of the simplification obtained by considering the reductive perturbation 
method, let us consider a Riemann wave:
\beq \label{i2}
\phi_{,t} + \lambda(\phi) \phi_{,x} = 0.
\eeq
When the wave function is small we may find the solution $\phi$ by a perturbation 
calculation. Let $\epsilon$ be a small parameter and let us expand the solution around the constant 
solution $\phi^{(0)}$:
\beq \nonumber
\phi = \phi^{(0)} + \epsilon \phi^{(1)} + O(\epsilon^2).
\eeq
Expanding in powers of $\epsilon$ we get from equation (\ref{i2}) the following results:
\bea 
\epsilon^0: && \, \phi^{(1)}_{,t} + \lambda_0 \phi^{(1)}_{,x} = 0, \nonumber \\ 
\epsilon^1: &&\, \phi^{(2)}_{,t} + \lambda_0 \phi^{(2)}_{,x} = - \lambda_{,\phi^{(0)}} \phi^{(1)} \phi^{(1)}_{,x},
\nonumber
\eea
with
$$
\lambda_0 \doteq \lambda(\phi^{(0)}), \qquad 
\lambda_{,\phi^{(0)}} \doteq \left( \frac{d\lambda}{d\phi} \right)_{\phi=\phi^{(0)}}.
$$
Introducing the new variables  $x' = x -\lambda_0 t$, $t' = \epsilon t$, we can rewrite the 
equation (\ref{i2}), up to the second order in $\epsilon$, as:
\bea \label{i6}
\phi^{(1)}_{,t'} + \lambda_{,\phi^{(0)}} \phi^{(1)} \phi^{(1)}_{,x'}=0.
\eea

If we consider a nonlinear dispersive system, like, for example, the Euler equation, 
instead of equation (\ref{i6}) we get:
\bea \label{i7}
\phi_{,t'} = \phi_{,x'x'x'} + 6  \phi  \phi_{,x'},
\eea
that is the Korteweg--de Vries (KdV) equation. 

The nonlinear system at the lowest order approximation can admit a solution given by 
monocromatic wave packets, i.e. $\phi^{(0)} = A\, {\rm{exp}}\, [kx - \omega(k) t]$.
Than it is reasonable to consider perturbations of such solution and  to turn the nonlinear 
system into a set of equations for the complex envelope of these packets. 
The characteristic packet size and wavelenght play the role of different scales for this 
system. 

Let us consider, for example, the  KdV equation (\ref{i7}) for a small amplitude field  $\phi$  of order $\epsilon$. 
The linear equations admits a monocromatic solution with dispersion relation     $\omega(k) = - k^3$.
Then the solution of the KdV equation can be written as
$$
\phi = \sum_{n=-\infty}^{+\infty} \epsilon^{\alpha_{| n |} }v_n(x',t') e^{in(kx-\omega(k)t)}, \qquad v_n^* = v_{-n}, 
$$
with
$$
x' =\epsilon(x + 3 k^2 t), \qquad t' = - 6 \epsilon^2 k t,
$$
$$
\alpha_0 = 1, \quad \alpha_n = n-1, \quad n \ge 1,
$$
and $v_1$ will satisfy 
the well known integrable nonlinear Sch\"rodinger (NLS) equation
\bea \label{i9}
v_{1,t'} + \frac{1}{2} v_{1,x'x'} - k^2 v_1 | v_1 |^2 = 0.
\eea

It is important to notice that these multi--scale expansions are structurally strong and 
can be applied to both integrable and non integrable systems.
Zakharov and Kuznetsov in the introduction of their article \cite{kz} say: 
{\it If the initial system is not integrable, the result can be both 
integrable and nonintegrable. But if we treat the integrable system properly, we 
again must get from it an integrable system.}

Calogero and Eckhaus \cite{calogero} used similar ideas starting from generic hyperbolic systems to 
prove in 1987 the necessary conditions for the integrability of nonlinear partial
differential equations (PDEs). 
Later Degasperis and Procesi \cite{dp} introduced the notion of {\it asymptotic 
integrability of order n } by requiring that the multi--scale expansion be verified 
up to order $n$. 

Also in the case of differential equations on a lattice, we would like to have a reliable reductive 
perturbative method which would produce reduced discrete systems. As the far field solution implies 
the introduction of a new variable which combines the continuous time with the discrete lattice, 
it is natural to get from a differential--difference equation by the reductive 
perturbation technique a continuous NLS 
equation (\ref{i9}).
Leon and Manna   \cite{leon} and later Levi and Heredero \cite{lh}  proposed a set of
 tools which allows to perform multiscale analysis for a  discrete evolution equation. 
These tools rely on the definition of a large grid scale via the comparison of the magnitude 
of related difference operators and on  the introduction of a slow varying condition for functions defined on the lattice. 
Their results, however, are not very promising as the reduced models are neither simpler nor as 
integrable as the original ones. 
Starting from an integrable model, like the Toda lattice, Leon and Manna \cite{leon} produce a 
non integrable differential--difference equation of the discrete 
NLS type. 
Levi and Heredero \cite{lh} from the integrable differential--difference NLS
equation got a non integrable system of differential--difference 
equations of KdV type.

\vspace{.2cm}

In the present paper we consider the case of nonliner partial difference equations (P$\Delta$Es).
To be able to carry out the discrete reductive perturbation technique,
in section \ref{sm} we introduce multiple lattice variables and give a definition of
slow varying functions on the lattice.
Section \ref{s3} is devoted to the application of the perturbative expansions introduced
to the case of a set of integrable and non integrable equations, i.e. the 
lattice modified Korteweg--de Vries (mKdV) equation, 
the Hietarinta equation, the lattice Volterra--Kac--Van Moerbeke (VKVM) equation and
a non integrable lattice KdV equation. Section \ref{conclusions} is devoted to some conclusive remarks.

%%%%%%%%%%%%%%%%%%%%%%%%%%%%%%%%%%%%%%%%%%%%%%%%%%%%%%%%%%%%%%%%%%%%%%%%%%%%%%%%%%%%%%%%%%%%%%%%%%%%%%%% 
%%%%%%%%%%%%%%%%%%%%%%%%%%%%%%%%%%%%%%%%%%%%%%%%%%%%%%%%%%%%%%%%%%%%%%%%%%%%%%%%%%%%%%%%%%%%%%%%%%%%%%%%
\section{Multiple--scales on the lattice  and functional variation on them} \label{sm}
%%%%%%%%%%%%%%%%%%%%%%%%%%%%%%%%%%%%%%%%%%%%%%%%%%%%%%%%%%%%%%%%%%%%%%%%%%%%%%%%%%%%%%%%%%%%%%%%%%%%%%%% 
%%%%%%%%%%%%%%%%%%%%%%%%%%%%%%%%%%%%%%%%%%%%%%%%%%%%%%%%%%%%%%%%%%%%%%%%%%%%%%%%%%%%%%%%%%%%%%%%%%%%%%%%

The aim of this  section is to fix the notation and to introduce the mathematical
formulae necessary  to reduce integrable and non integrable lattice equations in the
framework of the perturbative--reductive approach. In doing so we will partly follow \cite{levi}, 
trying to present a clearer and simpler derivation of all necessary formulae.

%%%%%%%%%%%%%%%%%%%%%%%%%%%%%%%%%%%%%%%%%%%%%%%%%%%%%%%%%%%%%%%%%%%%%%%%%%%%%%%%%%%%%%%%%%%%%%%%%%%%%%%% 
%%%%%%%%%%%%%%%%%%%%%%%%%%%%%%%%%%%%%%%%%%%%%%%%%%%%%%%%%%%%%%%%%%%%%%%%%%%%%%%%%%%%%%%%%%%%%%%%%%%%%%%%
\subsection{Slow varying variables on the lattice} \label{pp}
%%%%%%%%%%%%%%%%%%%%%%%%%%%%%%%%%%%%%%%%%%%%%%%%%%%%%%%%%%%%%%%%%%%%%%%%%%%%%%%%%%%%%%%%%%%%%%%%%%%%%%%% 
%%%%%%%%%%%%%%%%%%%%%%%%%%%%%%%%%%%%%%%%%%%%%%%%%%%%%%%%%%%%%%%%%%%%%%%%%%%%%%%%%%%%%%%%%%%%%%%%%%%%%%%%

Given a lattice defined by a constant lattice spacing $h$, we will
 denote by $n$ the running index of the points separated by $h$.
In correspondence with the lattice variable $n$,
we can introduce the real variables $x = hn$. 

We can define  on the same lattice a set of slow varying variables  by  introducing a small parameter $\epsilon=N^{-1}$ and requiring that
\bea \label{f2b}
n_j = \epsilon^j n.
\eea  
This is equivalent to sampling points from the original variables which are situated at a distance of $N^j h$ between them and then setting them on a lattice of spacing $h$. The corresponding slowly varying real variables $x_j$ are related to the variable $x$ by the equation $x_j = \epsilon^j x$.

%%%%%%%%%%%%%%%%%%%%%%%%%%%%%%%%%%%%%%%%%%%%%%%%%%%%%%%%%%%%%%%%%%%%%%%%%%%%%%%%%%%%%%%%%%%%%%%%%%%%%%%% 
%%%%%%%%%%%%%%%%%%%%%%%%%%%%%%%%%%%%%%%%%%%%%%%%%%%%%%%%%%%%%%%%%%%%%%%%%%%%%%%%%%%%%%%%%%%%%%%%%%%%%%%%
\subsection{Expansion of slowly varying functions.} \label{pp1}
%%%%%%%%%%%%%%%%%%%%%%%%%%%%%%%%%%%%%%%%%%%%%%%%%%%%%%%%%%%%%%%%%%%%%%%%%%%%%%%%%%%%%%%%%%%%%%%%%%%%%%%% 
%%%%%%%%%%%%%%%%%%%%%%%%%%%%%%%%%%%%%%%%%%%%%%%%%%%%%%%%%%%%%%%%%%%%%%%%%%%%%%%%%%%%%%%%%%%%%%%%%%%%%%%%

Let us
study the relation between functions living on the different lattices defined in section \ref{pp}.
We consider a function $f\doteq f_n$ defined on the points of a lattice of index $n$.
Let us assume
that $f_n = g_{n_1,n_2, \dots,n_K}$, i.e. $f$ depends on a finite
number $K$ of slow varying lattice variables $n_j$ $j=1,2,\ldots,K$ defined as in  (\ref{f2b}). 
We want to get explicit expressions for, say,  $f_{n+1}$ in terms of 
$g_{n_1, n_2, \ldots, n_K}$ evaluated on the points of the $n_1$, $n_2$, $\ldots, n_K$ lattices. 
At first let us consider the case,
studied in \cite{Jordan}, when we have only two different lattices,
i.e. $K = 1$. Using the results obtained in this case we will then
consider the case corresponding to $K = 2$. The general case will than be obvious.

\vspace{.5cm}

{\bf{I)}} $K=1$ ($f_n=g_{n_1}$). 
In this case we  use the following result presented in \cite{Jordan}:
\begin{equation} \label{difInt}
\Delta^k \, g_{n_1}\doteq \sum_{i=0}^{k}  (-1)^{k-i}
{k \choose i} g_{n_1+i} =\sum_{i=k}^\infty  {k!\over i!}P(i,k)\, \Delta^i  f_n.
\end{equation}
Here  the coefficients
$P(i,k)$ are given by
\bea \label{p1}
P(i,j)=\sum_{\alpha=j}^i
 \omega^\alpha S_{i}^{\alpha} \, \mathfrak{S}_{\alpha}^j,
\eea
where $\omega$ is the ratio of the increment in the lattice of variable $n$ 
with respect to that of variable $n_1$. In this case, taking into account equation (\ref{f2b}), 
$\omega=N$.
The coefficients $S_{i}^{\alpha}$ and $\mathfrak{S}_{\alpha}^j$ are the Stirling numbers of the
first and second kind respectively \cite{AS}. Formula (\ref{difInt}) allow us to express a difference of
order $k$ in the lattice of variable $n_1$ in terms of an infinite
number of differences on the lattice of variable $n$. The result (\ref{difInt}) can be inverted and we get:
\begin{equation} \label{difInt1}
\Delta^k \, f_{n}=\sum_{i=k}^\infty  {k!\over i!}Q(i,k)\, \Delta^i  g_{n_1},
\end{equation}
where  the coefficients
$Q(i,k)$ are given by (\ref{p1}) with $\omega=N^{-1}=\epsilon$.

To get from equations  (\ref{difInt}) and (\ref{difInt1}) 
a finite approximation of the variation of $g_{n_1}\doteq f_n$  we need  
to truncate the expansion in the r.h.s. 
by requiring a slow varying condition for the function $f_n$. 
Let us introduce the following definition:

\vspace{.1cm}

{\bf{Definition}}. {\it{The function $f_n$ is a slow varying function of
order $p$ iff $\Delta^{p+1} \, f_n=0$.}}

\vspace{.1cm}

From Definition (2.1) it follows that a slow varying function of 
order $p$ is a polynomial of degree $p$ in $n$. From equations  
(\ref{difInt}) and (\ref{difInt1}) we see that also the following statement holds:

\vspace{.1cm}

{\bf{Theorem}}. {\it{$f_n$ is a slow varying function of
order $p$ iff $\Delta^{p+1} \, g_{n_1}=0$, namely $g_{n_1}$
is of order $p$.}}

\vspace{.1cm}

Equation (\ref{difInt1}) provide us with  the formulae for
$f_{n+1}$ in terms of $g_{n_1}$  and its neighboring points in the case of slow varying functions of any order. Let us write down explicitly these expressions in the case of $g_{n_1}$ of order 1,2 and 3.

\begin{itemize}

\item $p=1$. The formula (\ref{difInt1}) reduces to
\beq \nonumber
\Delta f_{n} =\frac{1}{N}\Delta  g_{n_1},
\eeq
i.e. $f_{n+1}$ reads
\beq \nonumber
f_{n+1} = g_{n_1} + \frac{1}{N} (g_{n_1+1}- g_{n_1})+ O(N^{-2}).
\eeq

\item $p=2$. From equation (\ref{difInt1}) we get
\beq \nonumber
\Delta f_{n} = \frac{1}{N} \Delta  g_{n_1} + \frac{1-N}{2 N^2} 
\Delta^2 \, g_{n_1},
\eeq
and thus $f_{n+1}$ reads
\beq \label{p=22}
f_{n+1} = g_{n_1} + \frac{1}{2N} (-g_{n_1+2}+4 g_{n_1+1}-3 g_{n_1})+
\frac{1}{2N^2} (g_{n_1+2}-2 g_{n_1+1}+g_{n_1})+
O(N^{-3}).
\eeq

\item $p=3$. From equation (\ref{difInt1}) we get
\beq \nonumber
\Delta f_{n} = \frac{1}{N} \Delta g_{n_1} + 
\frac{1-N}{2 N^2} \Delta^2  g_{n_1}+ \frac{(1-N)(1-2N)}{6 N^3}\Delta^3  g_{n_1},
\eeq
and thus $f_{n+1}$ reads
\bea 
f_{n+1} &=& g_{n_1} + \frac{1}{6N} (2g_{n_1+3}-9g_{n_1+2}+13 g_{n_1+1}-6 g_{n_1})+  \nonumber \\
&& + \frac{1}{2N^2} (-g_{n_1+3}+4g_{n_1+2}-5g_{n_1+1}+2 g_{n_1})+ \nonumber \\
&& + \frac{1}{6N^3} (g_{n_1+3}-3g_{n_1+2}+3g_{n_1+1}- g_{n_1})+ O(N^{-4}). \nonumber
\eea

\end{itemize}

In the next sections we will consider mainly the reduction of 
integrable  discrete equations and we will be interested in obtaining from them
integrable discrete equations. It is known \cite{ly} that a scalar 
differential--difference equation can possess higher conservation laws and thus be integrable only if
it depends symmetrically on the discrete variable, i.e. if the discrete equation is invariant with respect to the
inversion of the lattice index. 
The results contained in (\ref{difInt1}) do not provide us with symmetric
formulas. To get symmetric formulas we  start from equation 
(\ref{difInt}) and take into account the
following remarks:
\begin{enumerate}
\item \label{r1} Formula (\ref{difInt}) holds also if  $h$ is negative;
\item \label{r2} For a slow varying function of order $p$, we have $\Delta^p  f_n = \Delta^p  f_{n +\ell}$,
for all $\ell \in \mathbb{Z}$.
\end{enumerate}
When  $f_n$ is a slow varying function of odd order we are not able
to construct completely symmetric derivatives using
just an odd number of points centered around the $n_1$ point and thus $f_{n \pm 1}$ can never be expressed 
in a symmetric form.

Using the above remarks we can construct the symmetric version of (\ref{p=22}). 
From (\ref{difInt}) we get:
\beq \label{p21}
g_{n_1+1} = g_{n_1} + N { \Delta}  f_n + \half N (N-1) { \Delta}^2  f_n, 
\eeq
where ${ \Delta}^2 f_n= f_{n + 1} - 2 f_n + f_{n - 1}$ thanks to the
remark \ref{r2}. Using the remark \ref{r1} we can also write:
\beq \label{p22}
g_{n_1-1} = g_{n_1} +N { \Delta_{-1}}  f_n + \half N (N-1) { \Delta}^2  f_n, 
\eeq
where ${ \Delta_{-1}} \, f_n \doteq f_{n-1} - f_n$. From equations 
(\ref{p21}) and (\ref{p22}) we obtain the following form
for $f_{n + 1}$:
\beq \label{p23}
f_{n + 1} =
g_{n_1}+  \frac{1}{2 N}(g_{n_1+1} - g_{n_1-1}) + 
\frac{1}{2 N^2}(g_{n_1+1}- 2 g_{n_1}+ g_{n_1-1})+
O(N^{-3}).
\eeq

\vspace{.5cm}

{\bf{II)}} $K=2$ ($f_n=g_{n_1,n_2}$). The derivation of the formulae in this case is done in 
the same spirit as for the symmetric expansion presented above, 
see equation (\ref{p23}). Let us just consider the case when $p=2$, 
as this is the lowest value of $p$ for which we can consider 
$f_n$ as a function of the two scales $n_1$ and $n_2$.  From equation
(\ref{p21}) we get:
\bea \label{p24}
&& g_{n_1+1,n_2} = g_{n_1,n_2} + N \Delta_1 f_{n,n} + \half N ( N-1) \Delta_1^2 f_{n,n}, \\ \label{p25}
&& g_{n_1,n_2+1} = g_{n_1,n_2} + N^2 \Delta_2 f_{n,n} + \half N^2 ( N^2-1) \Delta_2^2 f_{n,n}.
\eea
Here the symbols $\Delta_1$ and $\Delta_2$ 
denote difference operators which acts on the first and respectively on the second index 
of the function $f_{n,n} \doteq g_{n_1,n_2}$, e.g. $\Delta_1 f_{n,n} \doteq f_{n+1,n} - f_{n,n}$
and $\Delta_2 f_{n,n} \doteq f_{n,n+1} - f_{n,n}$.

Let us now consider a function $g_{n_1,n_2}$ 
where one shifts both indices by $1$. From equation (\ref{p24}), taking into account that, from equation (\ref{f2b}), for example, $g_{n_1+1,n_2} = f_{n+N,n}$,  one has:
\bea \label{p26}
g_{n_1+1,n_2+1} = g_{n_1,n_2+1} + N \Delta_1 f_{n,n+N^2} + \half N ( N-1) \Delta_1^2 f_{n,n+N^2}, 
\eea
and using the result (\ref{p25}) we can write equation (\ref{p26}) as
\bea 
g_{n_1+1,n_2+1} &=& g_{n_1,n_2} + N^2 \Delta_2 f_{n,n} + 
\half N^2 ( N^2-1) \Delta_2^2 f_{n,n} + \label{p27} \\ 
&& + N  \Delta_1 \left [ f_{n,n} + N^2 \Delta_2 f_{n,n} + 
\half N^2 ( N^2-1) \Delta_2^2 f_{n,n} \right ] +\nonumber  \\ 
&& + \half N ( N-1) \Delta_1^2 \left [ f_{n,n} + 
N^2 \Delta_2 f_{n,n} + \half N^2 ( N^2-1) \Delta_2^2 f_{n,n} \right ] = \nonumber \\ 
&=&  g_{n_1,n_2} + N^2 \Delta_2 f_{n,n} + \half N^2 ( N^2-1) \Delta_2^2 f_{n,n} + \nonumber \\ 
&& + N  \Delta_1  f_{n,n} + N^3 \Delta_1 \Delta_2 f_{n,n} + 
\half N^3 ( N^2-1) \Delta_1 \Delta_2^2 f_{n,n} + \nonumber \\ 
&& + \half N ( N-1) \Delta_1^2  f_{n,n} + 
N^3 ( N-1) \Delta_1^2 \Delta_2 f_{n,n} + \frac{1}{4} N^3 ( N^2-1)( N-1) \Delta_1^2 \Delta_2^2 f_{n,n}.
\nonumber
\eea
As, using the second remark, the second difference of $f_{n,n}$ depends just on 
its nearest neighboring points, the right hand side of equation (\ref{p27}) depends, 
apart from $f_{n,n}=g_{n_1,n_2}$, on 
$f_{n,n+1}, \, f_{n,n-1}, \, f_{n+1,n}, \, f_{n-1,n}$, $ f_{n+1,n+1}, \, f_{n+1,n-1}, \, f_{n-1,n+1}, $ 
and $f_{n-1,n-1}$, i.e. 8 unknowns. Starting from equations (\ref{p24}), (\ref{p25}) and (\ref{p27}) 
we can write down 8 equations, using the first remark, 
which define $g_{n_1+1,n_2}, \, g_{n_1-1,n_2}, \, g_{n_1,n_2+1}$, $g_{n_1,n_2-1}, \, g_{n_1+1,n_2+1}$, 
$g_{n_1+1,n_2-1}, \, g_{n_1-1,n_2+1},$ and $g_{n_1-1,n_2-1}$ in terms of  the 
functions $f_{n+i,n+j}$ with $(i,j)=0,\pm1$. Inverting this system of 
equations we get $f_{n \pm 1}$ in term of $g_{n_1,n_2}$ and its shifted values:
\bea 
f_{n \pm 1} &=& g_{n_1,n_2} \pm  \frac{1}{ 2 N} (g_{n_1+1,n_2}   - g_{n_1-1,n_2}) +
\frac{1}{2 N^2} (g_{n_1+1,n_2} - 2 g_{n_1,n_2} + g_{n_1-1,n_2}) 
+ \nonumber  \\ 
&& \pm  \frac{1}{2 N^2}(g_{n_1,n_2+1} - g_{n_1,n_2-1})+ \nonumber \\
&& +  \frac{1}{4 N^3} 
( g_{n_1+1,n_2+1} - g_{n_1-1,n_2+1} - g_{n_1+1,n_2-1} + g_{n_1-1,n_2-1}) + 
O(N^{-4}).  \label{p27b}
\eea
It is worthwhile to notice that the two lowest order (in $N^{-1}$) terms of the expansion (\ref{p27b}) 
are just the sum of the first symmetric differences of $g_{n_1}$ and $g_{n_2}$. 
Thus in the continuous limit, when we divide by $h$ and send $h$ to zero in 
such a way that $x=hn$, $x_1=hn_1$ and $x_2=hn_2$ be finite, we will have 
\bea \nonumber
f_{,x} = \epsilon g_{x_1} + \epsilon^2 g_{x_2}.
\eea 
Extra terms appear at the order $N^{-3}$ and contain shifts in both $n_1$ and $n_2$.

When $f_n$ is a slow varying function of order 2 in $n_1$  it can also be 
 of order 1  in $n_2$. In such a case equation (\ref{p25}) is given by
\bea \label{p28}
g_{n_1,n_2+1} = g_{n_1,n_2} + N \Delta_2 f_{n,n}.
\eea
Starting from equations (\ref{p24}), (\ref{p28})  and a modified (\ref{p27}) we can get a set of 8 equations which allows us 
to get $f_{n \pm 1}$ in terms of $g_{n_1,n_2}$ and its shifted values.  In such a case 
 $f_{n \pm 1}$ reads
\bea
f_{n \pm 1} &=& g_{n_1,n_2} \pm \frac{1}{ 2 N} (g_{n_1+1,n_2}  - g_{n_1-1,n_2})
+ \frac{1}{ N^2} (g_{n_1,n_2 \pm 1} - g_{n_1,n_2} )  + \nonumber \\
&& + \frac{1}{2 N^2}(g_{n_1+1,n_2} - 2 g_{n_1,n_2} + g_{n_1-1,n_2})+
O(N^{-3}). \label{h16} 
\eea

It is possible to introduce two parameters in the definition of $n_1$, $n_2$ in terms of $n$. Let us
define
$$
n_1 = \frac{n M_1}{N}, \qquad n_2 = \frac{n M_2}{N^2},
$$
where $M_1$ and $M_2$ are divisors of $N$ and $N^2$ so that $n_1$ and $n_2$  are integers numbers. 
In such a case equation (\ref{p27b}) reads
\bea 
f_{n \pm 1} &=& g_{n_1,n_2} \pm  \frac{M_1}{ 2 N} (g_{n_1+1,n_2}   - g_{n_1-1,n_2}) +
\frac{M_1^2}{2 N^2} (g_{n_1+1,n_2} - 2 g_{n_1,n_2} + g_{n_1-1,n_2}) 
+ \nonumber  \\ 
&& \pm  \frac{M_2}{2 N^2}(g_{n_1,n_2+1} - g_{n_1,n_2-1})+ \nonumber \\
&& +  \frac{M_1 M_2}{4 N^3} 
( g_{n_1+1,n_2+1} - g_{n_1-1,n_2+1} - g_{n_1+1,n_2-1} + g_{n_1-1,n_2-1}) + 
O(N^{-4})  \label{h16a}
\eea
and equation (\ref{h16}) accordingly.

When we consider partial difference equations we have more than one independent variable.  Let us consider  the case of two independent lattices and a function $f_{n,m}$ defined on them. As the two lattices are independent the formulae 
presented above apply independently 
on each of the lattice variables. So, for instance, the variation $f_{n+1,m}$ 
when the function $f_{n,m}$ is a slowly varying function of order 2 of a lattice variable $n_1$ reads
\bea \label{h16b}
f_{n+1,m} =g_{n_1,m} +  \frac{1}{2 N} (g_{n_1+1,m} - g_{n_1-1,m}) + 
\frac{1}{2 N^2}(g_{n_1+1,m} - 2 g_{n_1,m}+ g_{n_1-1,m}) +
O(N^{-3}).
\eea

A slightly less obvious situation appears when we consider $f_{n+1,m+1}$, as new terms will appear. 
We consider here just the case we will need later when 
\bea \label{p31}
n_1 = \frac{M_1 n}{N}, \qquad m_1 = \frac{M_2 m}{N}, \qquad m_2 = \frac{n}{N^2}.
\eea
If $f_{n,m}$ is a slow varying function of first order in $m_2$ and of second order in both 
$n_1$ and $m_1$, from equations (\ref{h16a}) and (\ref{h16b}) the variation $f_{n+1,m+1}$ reads
\bea 
f_{n+1,m+1} &= & g_{n_1,m_1,m_2} + \frac{M_1}{ 2 N} (g_{n_1+1,m_1,m_2}   - g_{n_1-1,m_1,m_2}) + 
\frac{M_2}{ 2 N} (g_{n_1,m_1+1,m_2}   - g_{n_1,m_1-1,m_2}) + \nonumber \\ 
&& + \frac{M_1^2}{ 2 N^2} (g_{n_1+1,m_1,m_2}   + g_{n_1-1,m_1,m_2} - 2 g_{n_1,m_1,m_2}) + \nonumber \\
&& + \frac{M_2^2}{ 2 N^2} (g_{n_1,m_1+1,m_2}   + g_{n_1,m_1-1,m_2} - 2 g_{n_1,m_1,m_2}) + \nonumber \\ 
&& + \frac{M_1 M_2}{ 4 N^2} (g_{n_1+1,m_1+1,m_2}   + g_{n_1-1,m_1-1,m_2} - g_{n_1+1,m_1-1,m_2} - 
g_{n_1-1,m_1+1,m_2}) + \nonumber \\
&& + \frac{1}{  N^2}
(g_{n_1,m_1,m_2+1} - g_{n_1,m_1,m_2} )  +O(N^{-3}). \nonumber
\eea

%%%%%%%%%%%%%%%%%%%%%%%%%%%%%%%%%%%%%%%%%%%%%%%%%%%%%%%%%%%%%%%%%%%%%%%%%%%%%%%%%%%%%%%%%%%%%
\section{ Multiscale reduction of nonlinear partial difference equations} \label{s3}
%%%%%%%%%%%%%%%%%%%%%%%%%%%%%%%%%%%%%%%%%%%%%%%%%%%%%%%%%%%%%%%%%%%%%%%%%%%%%%%%%%%%%%%%%%%%%%
In the following we will apply the formulae obtained in section 2 to some well known partial difference equations. Some of those are known to have a Lax pair and are  associated to integrable partial differential equations. Others are concocted so as to have a real dispersion relation but with no particular reason why they should be integrable. The integrable equations we will consider here, the lattice modified KdV (mKdV), presented in section \ref{mkdv}, the Hietarinta equation, presented in section \ref{hieta} and the lattice Volterra--Kac--Van Moerbeke (VKVM) equation, presented in section \ref{vkvm}, are  defined on four lattice points and are P$\Delta$Es {\sl consistent around a cube} \cite{hietarinta}.  From this property one can derive their Lax equation. 

The lattice mKdV is an integrable equation of the same class of the lattice potential KdV and KdV \cite{calogero} and it 
possesses a Lax pair \cite{frank1}.  As from KdV we get by multiscale reduction the NLS \cite{kz}, 
the same we may expect here. To get an integrable discrete equation we expect  
a resulting discrete equation which is somehow symmetric. At least when  $h_t \rightarrow 0$ 
with $t =m \, h_t$ the differential difference equation we obtain must be symmetric in terms of 
the inversion of $n_j$, i.e. if it contains  $n_{j+k}$ it will contain also  $n_{j-k}$. 

The non integrable KdV equation presented in section \ref{nonintegrable} is obtained by a straightforward  discretization using the symmetric representation of the derivatives so as to get a real dispersion relation.  

In all the cases considered we will expand the solution of the nonlinear lattice equation around a wave solution of the linear part. In doing so we require that the wave solution be always bounded so that a perturbative expansion with slowly variable coefficients is meaningful. This can be always achieved if the dispersion relation is real. This is always true if the equation can be rewritten in terms of symmetric derivatives.
Moreover to get a meaningful reduction we need to have a non trivial nonlinear dispersion relation. 

\subsection{Reduction of the lattice mKdV} \label{mkdv}
%%%%%%%%%%%%%%%%%%%%%%%%%%%%%%%%%%%%%%%%%%%%%%%%%%%%%%%%%%%%%%%%%%%%%%%%%%%%%%%%%%%%%%%%%%%%%%%%%%%%%%%% 
%%%%%%%%%%%%%%%%%%%%%%%%%%%%%%%%%%%%%%%%%%%%%%%%%%%%%%%%%%%%%%%%%%%%%%%%%%%%%%%%%%%%%%%%%%%%%%%%%%%%%%%%

The discrete analogue of the modified Korteweg--de Vries (mKdV) equation is given by the
following nonlinear P$\Delta$E \cite{frank1}:
\beq \label{3.1}
p \, ( u_{n,m} \, u_{n,m+1} - u_{n+1,m} \, u_{n+1,m+1} ) - 
q  \, ( u_{n,m} \, u_{n+1,m} - u_{n,m+1} \, u_{n+1,m+1} )=0.
\eeq
This equation involves just four points which lay on two  
orthogonal infinite lattices and are the vertices of an elementary square.
In equation (\ref{3.1})  $u_{n,m}$ is the dynamical field (real) variable at site 
$(m,n) \in \mathbb{Z} \times \mathbb{Z}$ and 
 $p,q \in \mathbb{R}$ are the lattice parameters. These are assumed different from zero and will go to zero in the continuous limit so as to get the continuous mKdV.

Carrying out the change of variable
$
u_{n,m} \mapsto
1 + u_{n,m},
$ 
one can separate the linear and nonlinear parts of equation (\ref{3.1}):
\bea
&& p \,  
( u_{n,m} + u_{n,m+1} - u_{n+1,m}- u_{n+1,m+1} )- 
q\,  ( u_{n,m} +u_{n+1,m} - u_{n,m+1}- u_{n+1,m+1} )=  \nonumber \\ 
&=& q \,  ( u_{n,m} \, u_{n+1,m} - u_{n,m+1}\,  u_{n+1,m+1} )
- p \, ( u_{n,m}\,  u_{n,m+1} - u_{n+1,m} \, u_{n+1,m+1} ) . \label{kk} 
\eea

Let us consider the linear part of equation (\ref{kk}), namely 
\beq \label{3.2a}
p \, ( u_{n,m}+ u_{n,m+1} - u_{n+1,m}- u_{n+1,m+1})- 
q \,  ( u_{n,m} +u_{n+1,m} - u_{n,m+1}- u_{n+1,m+1} )= 0.
\eeq
Given any initial condition $u_{n,0}$ the general solution of equation (\ref{3.2a}) is given by
\beq \label{3.7}
u_{n,m} = \frac{1}{2 \pi {\rm {i}} } \sum_{j=-\infty}^{\infty} u_{j,0} \oint_{|z|=1} 
\left[ \frac{(p-q) - (p+q)z}{(p-q) z - (p+q)} \right]^m   ~ z^{n-j-1} dz.
\eeq
Equation (\ref{3.7}) can be rewritten 
in a more natural way (from the continuous point of view) by defining
\beq \label{zO}
z \doteq e^{{\rm {i}} \, k},  \qquad  \Omega \doteq  e^{-{\rm {i}}\,   \omega} = 
\frac{(p-q) - (p+q)z}{(p-q) z - (p+q)} .
\eeq
In such a case  the solution 
(\ref{3.7}) is written as a superposition of linear waves
\beq \label{3.8d}
E_{n,m} = e^{{\rm {i}} [k\, n-\omega(k) \, m]} = z^n \, \Omega^m.
\eeq
The  dispersion relation for these linear waves is given by 
\beq \label{dr}
\omega = -2 \arctan \left[ \frac{p}{q} \tan \left( \frac{k}{2} \right)  \right],
\eeq 
the same as for the lattice potential KdV (pKdV) equation \cite{levi}. 
From equations (\ref{zO}) and (\ref{dr}), 
by differentiation with respect to $k$, we get the group velocity $\omega_{,k}$:
\beq \label{l1}
\omega_{,k} = \, \frac{4 \, p \, q \, z}{[(p-q)z - (p+q)] \, [(p+q)z - (p-q)]} 
=\frac{2 \, p \, q}{ p^2 + q^2 - (p^2 - q^2) \cos k}.
\eeq

The linear part of the P$\Delta$E (\ref{kk}), i.e. equation (\ref{3.2a}), is solved in terms of harmonics
(\ref{3.8d}) if $\omega$ is given by (\ref{dr}). The nonlinearity will couple the harmonics. This suggests  to
look for solutions of the P$\Delta$E (\ref{kk}) written as a combination of modulated waves:
\beq \label{3.9}
u_{n,m} = \sum_{s=0}^{\infty} \epsilon^{\beta_s}
\psi^{(s)}_{n,m}\,  (E_{n,m})^s +
\sum_{s=1}^{\infty} \epsilon^{\beta_s}
{\bar \psi}^{(s)}_{n,m} \, ({\bar E}_{n,m})^s, 
\eeq
where the functions $\psi^{(s)}_{n,m}$ are  slowly varying functions on the lattice, 
i.e. $\psi^{(s)}_{n,m}=\psi^{(s)}_{n_1,m_1,m_2}$  and $\epsilon^{\gamma} = N^{-1}$. 
By  $\bar b$ we mean the complex conjugate of a complex quantity $b$ so that, for example, 
${\bar E}_{n,m} = (E_{n,m})^{-1}$. 
The positive numbers $\beta_s$ are to be determined in such a way that :
\begin{enumerate}
\item $\beta_1 \leq \beta_s \; \forall \,  s=0,2,3,\dots,\infty$.
In general it is possible to set $\beta_1=1$.
\item In the equation for $\psi^{(1)}_{n,m}=\psi_{n,m}$, we require that the lowest order  nonlinear terms should match  the slow 
time derivative of the linear part after having solved all linear equations. This will provide a relation between $\gamma$ and the $\beta_s$.
\end{enumerate}
The fact that the second summation in equation (\ref{3.9}) starts from $s=1$ and contains the complex conjugates of the terms of the first summation is due to the reality condition for the solutions of the P$\Delta$E (\ref{3.1}).

After introducing the expansion (\ref{3.9}) 
in the P$\Delta$E (\ref{kk}) and analizing the coefficients of the various harmonics $(E_{n,m})^s$ 
for $s=1$, $s=2$ and $s=0$ (as, assuming that $\beta_s$ increases with $s$,  the nonlinear terms
will depend only on the lowest $s$ terms) we came to the conclusion that we can choose 
\bea \label{3.10b}
\gamma=1,  \quad \beta_0 = 2,  \quad \beta_s = s, \quad s \ge 1. 
\eea
The discrete slow varying variables $n_1$, $m_1$ and $m_2$ are defined in terms of $n$ and $m$ by
equation (\ref{p31}).

Having fixed the constants $\beta_s$ according to equation (\ref{3.10b}) we can introduce the ansatz (\ref{3.9}) 
into equation (\ref{kk}) and  get the determining equations. 

For $s=1$
we get, at lowest order in $\epsilon$, 
\beq \nonumber
\psi_{n_1,m_1,m_2}  [ (q-p) (1 - z \, \Omega) - (p + q) ( \Omega - z) ] = 0,
\eeq
which is identically solved by the dispersion relation (\ref{zO}). 

At $\epsilon^2$ we get the linear equation
\bea 
&& M_1 \,z \, [(p-q)\Omega  + (p+q) ] \,
(\psi_{n_1+1,m_1,m_2} - \psi_{n_1-1,m_1,m_2} )+  \nonumber \\ 
&+&M_2 \,\Omega \, [ (p-q)   z  - (p+q)   ] \, (  \psi_{n_1,m_1+1,m_2} - \psi_{n_1,m_1-1,m_2} )
= 0, \label{3.11d}
\eea
whose solution is given by 
\beq \nonumber
\psi_{n_1,m_1,m_2} = \phi_{n_2,m_2}, \qquad n_2 = n_1 - m_1.
\eeq
provided that the integers $M_1$ and $M_2$ are choosen as
\bea \label{m1m2}
M_1 = S \, \Omega \, [ (p-q) z  - (p+q) ], \qquad
M_2 = S \, z \, [ (p-q) \Omega  + (p+q) ], 
\eea
where $S \in \mathbb{C}$ is a constant; $S$ cannot be completely arbitrary since 
$M_1$ and $M_2$ are to be integer numbers. We will show in Appendix A how it is possible to choose the complex 
constant $S$ in such a way that $M_1$ and $M_2$ are
in fact integer numbers as required by equations (\ref{p31}).
Substituting the expression of $\Omega$ given in (\ref{zO})
into equation (\ref{m1m2}), we can rewrite $M_1$ and $M_2$ as:
\bea \label{l2}
M_1 = - S \,  [ (p+q) z  - (p-q) ], \qquad
M_2 = \frac{ 4 \, p \,q S \, z}{  [ (p+q) -  z (p-q) ]}.
\eea 
From equations (\ref{l1}) and (\ref{l2}) we get:
\bea \label{l3}
\omega_{,k} = \frac{M_2}{M_1},
\eea
i.e. the ratio $M_2 / M_1$ is the group velocity. 
As $M_1$ and $M_2$ are integers, it follows that not all values of $k$ are admissible 
as $\omega_{,k} \in \mathbb{Q}$.
Let us notice that also $n_2 = n_1 + m_1$ solves equation (\ref{3.11d}) by an appropriate choice of  
$M_1$ and $M_2$.

At $\epsilon^3$ we get a nonlinear equation for $\phi_{n_2,m_2}$ which
depends on $\psi^{(2)}_{n_2,m_2}$:
\bea \label{yyk}
&& \phi_{n_2,m_2+1} - \phi_{n_2,m_2} +  
   c_1 \, ( \phi_{n_2+2,m_2} + \phi_{n_2-2,m_2} - 2 \, \phi_{n_2,m_2} )  + \nonumber  \\ 
&+& c_2 \,( \phi_{n_2+1,m_2} +
\phi_{n_2-1,m_2} - 2 \, \phi_{n_2,m_2} ) +c_3  \, \psi^{(2)}_{n_2,m_2} \, {\bar \phi}_{n_2,m_2} = 0,
\eea
where
\bea 
c_1 &=& p\, q\left( p-q \right) S^{2} \, z^2 \, \frac{(p-q) - (p+q)z}
{ \left[ (p-q)z - (p+q) \right]^2}, \nonumber \\
c_2 &=& 2 \, p\, q\left( p-q \right)  S^{2} \, z \, 
\frac{(p+q)(1+z^2)-2\,(p-q)z}
{ \left[ (p-q)z - (p+q) \right]^2}, \label{yyka} \\
c_3 &=& 
\frac{2\, p\, q\left( p^2-q^2 \right) (1-z^2)^3 }
{z  \left[ (p-q)z - (p+q) \right]^2 \left[ (p+q)z -(p-q) \right]^2}.
\nonumber
\eea
Using the form of the complex constant $S$ obtained in  Appendix A, the coefficients (\ref{yyka})  read
\bea 
c_1 &=& -\frac{M_2^2 \, (p-q)}{16 \, p\,q}\,
\left[ \left( p+q \right)( \cos k +{\rm{i}}\, \sin k) -(p-q)   \right]   , \nonumber \\
c_2 &=& \frac{M_2^2 \, (p-q)}{4 \, p\,q}\,
\left[ \left( p+q \right)\cos k -(p-q)
\right]   , \label{ccc}  \\
c_3 &=& {\rm{i}}\,   \frac{2\,p\,q\,(p^2-q^2)\sin^3 k}{[(p^2+q^2)-(p^2-q^2) \cos k]^2}.
\nonumber
\eea
The coefficients (\ref{ccc}) depend on  the integer constant $M_2$. The integer $M_1$
is then written out in terms of $M_2$ and reads 
$$
M_1=M_2
\frac{p^2+q^2 - (p^2-q^2) \cos k}{2\, p\,q},
$$
so that not all values of $k$ are admissible as $M_1$ must be also an integer. See Appendix A for details.

The lowest order equations for the harmonic $s=2$ appear at $\epsilon^2$ and give
$$
\psi^{(2)}_{n_2,m_2} = \half (\phi_{n_2,m_2})^2.
$$
It is easy to see that the choice (\ref{3.10b}) implies that the coefficients of all other harmonics are expressed in terms of $\phi_{n_2,m_2}$ and ${\bar \phi}_{n_2,m_2}$.

Taking these results into account the nonlinear equation P$\Delta$E (\ref{yyk}) for $\phi_{n_2,m_2}$ reads:
\bea \label{yyy}
&& \; \; \; \;  {\rm {i}} \, (\phi_{n_2,m_2+1} - \phi_{n_2,m_2})
= C_1 \, (\phi_{n_2+2,m_2} + \phi_{n_2-2,m_2} - 2 \, \phi_{n_2,m_2}  ) + \nonumber \\ 
&&+ \, C_2 \, ( \phi_{n_2+1,m_2} + \phi_{n_2-1,m_2} - 2 \,\phi_{n_2,m_2} ) +
C_3 \, \phi_{n_2,m_2}\, |\phi_{n_2,m_2}|^2,
\eea
where $C_i= -  {\rm {i}} \, c_i$, $i=1,2$, $C_3= - {\rm {i}} \, c_3 /2$, 
and the coefficients $c_i$'s are given by equation
(\ref{ccc}). It is easy to see that $C_3$ is a real coefficient.

The P$\Delta$E (\ref{yyy}) is a {\it completely discrete  and local} NLS equation depending on
the first and second  neighboring lattice points. 
At difference from the Ablowitz and Ladik \cite{al} discrete NLS, the nonlinear term 
in (\ref{yyy}) is completely local.
The P$\Delta$E (\ref{yyy}) has a natural semi--continuous limit when $m_2 \rightarrow \infty$ as
$H_2 \rightarrow 0$ in such a way that $t_2= m_2\, H_2 \in \mathbb{R}$ is finite. 
Setting $n_2 \doteq n$ and $t_2 \doteq t$ one gets the following
nonlinear differential--difference equation:
\beq \label{yyyy}
{\rm {i}} \, \frac{\partial \phi_n}{\partial t}
= C_1 \, (\phi_{n+2} + \phi_{n-2} - 2 \,\phi_{n}  ) 
+ \, C_2 \, ( \phi_{n+1} + \phi_{n-1} - 2 \,\phi_{n} ) +
C_3 \, \phi_{n}\, |\phi_{n}|^2.
\eeq
The continuous limit of the  P$\Delta$E (\ref{yyy}) is obtained if we consider in equation (\ref{yyyy}) the limit
$n \rightarrow \infty$ as
$H_1 \rightarrow 0$ in such a way that $x= n\, H_1 \in \mathbb{R}$ is finite. The resulting
NLS equation reads
\beq \label{yyyy1}
{\rm {i}} \, \phi_{,t}=
(4\,C_1 +C_2) \phi_{,xx}+ C_3 \, \phi \,  |\phi|^2,
\eeq
where
\bea \label{yyyy2}
4\,C_1 +C_2 = -\frac{M^2_2\, (p^2-q^2) \sin k}{4\,p\,q}.
\eea
As the coefficient (\ref{yyyy2}) is real, equation
(\ref{yyyy1}) is just the well known integrable NLS equation.

%%%%%%%%%%%%%%%%%%%%%%%%%%%%%%%%%%%%%%%%%%%%%%%%%%%%%%%%%%%%%%%%%%%%%%%%%%%%%%%%%%%%%%%%%%%%%%%%%%%%%%%% 
%%%%%%%%%%%%%%%%%%%%%%%%%%%%%%%%%%%%%%%%%%%%%%%%%%%%%%%%%%%%%%%%%%%%%%%%%%%%%%%%%%%%%%%%%%%%%%%%%%%%%%%%
\subsection{Reduction of the Hietarinta equation} \label{hieta}
%%%%%%%%%%%%%%%%%%%%%%%%%%%%%%%%%%%%%%%%%%%%%%%%%%%%%%%%%%%%%%%%%%%%%%%%%%%%%%%%%%%%%%%%%%%%%%%%%%%%%%%% 
%%%%%%%%%%%%%%%%%%%%%%%%%%%%%%%%%%%%%%%%%%%%%%%%%%%%%%%%%%%%%%%%%%%%%%%%%%%%%%%%%%%%%%%%%%%%%%%%%%%%%%%%

In \cite{hietarinta} 
Hietarinta introduces a new {\it consistent around a cube} P$\Delta$E 
\beq \label{hiej}
\frac{u_{n,m}+e_2}{u_{n,m}+e_1}
\frac{u_{n+1,m+1}+o_2}{u_{n+1,m+1}+o_1}-
\frac{u_{n+1,m}+e_2}{u_{n+1,m}+o_1}
\frac{u_{n,m+1}+o_2}{u_{n,m+1}+e_1}=0,
\eeq
where the four constants $e_i,o_i \in \mathbb{R}$, $ 1\leq i \leq 2$, are lattice parameters.

By a direct calculation one can separate the linear and the nonlinear parts of the
equation (\ref{hiej}):
\bea \label{hie2}
&&  \quad  o_1 \, o_2 (e_1 -e_2) u_{n,m}+ e_1 \, e_2 (o_1 -o_2)u_{n+1,m+1} + \nonumber \\
&& + \; e_1 \, o_2 ( e_2 - o_1) u_{n+1,m} + \, e_2 \, o_1 ( o_2 - e_1) u_{n,m+1} =  \nonumber \\
&& = [(o_2-e_1) u_{n+1,m} + (e_2-o_1) u_{n,m+1}  ] u_{n,m} \, u_{n+1,m+1} + \nonumber \\
&& +\;   [(o_1-o_2) u_{n,m} + (e_1-e_2) u_{n+1,m+1}  ] u_{n+1,m} \, u_{n,m+1} + \nonumber \\
&& +\; [o_1(e_2-o_2) u_{n,m+1} + o_2 (o_1-e_1) u_{n+1,m}  ] \, u_{n,m} + \nonumber \\
&& + \;[e_2(e_1-o_1) u_{n,m+1} + e_1 (o_2-e_2) u_{n+1,m}  ] \, u_{n+1,m+1}+ \nonumber \\
&& + \; (o_2 \, e_2 - o_1 \, e_1) (u_{n,m} \, u_{n+1,m+1} - u_{n+1,m} \, u_{n,m+1}).   
\eea

Let us now solve the linear part of the P$\Delta$E (\ref{hie2}):
\bea \label{hieL}
&&  \quad  o_1 \, o_2 (e_1 -e_2) u_{n,m}+ e_1 \, e_2 (o_1 -o_2)u_{n+1,m+1} + \nonumber \\
&& + \; e_1 \, o_1 ( e_2 - o_2) u_{n+1,m} + \, e_2 \, o_2 ( o_1 - e_1) u_{n,m+1} =0.
\eea
Defining 
\beq \nonumber
z \doteq e^{{\rm {i}} \, k},  \qquad  \Omega \doteq  e^{-{\rm {i}} \,  \omega} = 
\frac{o_2 \, [ e_1 ( e_2 - o_1 ) z+ o_1 ( e_1 - e_2)] }
{e_2 \, [ e_1 ( o_2 - o_1 ) z+ o_1 ( e_1 - o_2)] },
\eeq
the (complex) dispersion relation for these linear waves is given by 
\beq \nonumber
\omega = 2 \arctan \left\{
\frac{{\rm{i}} \, e_1 \, o_1 ( o_2 - e_2) \tan \left( k/2\right)} 
{[o_1 \, o_2 \, (e_1 - e_2)+ e_1 \, e_2 \, (o_1 - o_2)] \tan \left( k/2\right)
+{\rm{i}} \,  e_2 \, o_2 ( e_1 - o_1)}
\right\}.
\eeq
The dispersion relation is a real function of $k$ if the following condition
holds:
\beq \label{rch}
o_1 \, o_2 \, (e_1 - e_2)+ e_1 \, e_2 \, (o_1 - o_2)=0.
\eeq
To give a meaning to the expansion (\ref{3.9}) we have to require that the dispersion relation $\omega(k)$ be a real function for $k$ real. 
From equation (\ref{rch})  we can 
write $o_2$ in terms of $o_1,e_1,e_2$, and get:  
\beq \label{newO}
\Omega \doteq  e^{-{\rm {i}} \,  \omega} = 
\frac{e_1(e_2-o_1)z +o_1 (e_1-e_2)}
{o_1(e_1-e_2)z +e_1 (e_2-o_1)},
\eeq
so that the real dispersion relation reads
\beq \label{real}
\omega = 2 \arctan \left[
\frac{e_2(e_1+o_1) -2 \, e_1 \, o_1} 
{e_2 ( o_1 - e_1)} \tan \left( \frac{k}{2}\right)
\right].
\eeq
From equation (\ref{newO}), by differentiation with respect to $k$, we get the real group velocity $\omega_{,k}$
\beq \label{gvh}
\omega_{,k} = 
\frac{e_2( o_1 - e_1 )[2\,  e_1 \, o_1 -e_2(e_1+o_1)]\, z}
{[e_1(o_1-e_2)z +o_1 (e_2-e_1)][o_1(e_1-e_2)z +e_1 (e_2-o_1)] }.
\eeq 

The P$\Delta$E (\ref{hieL}) has a bounded wave solution given
by equation (\ref{3.8d}), where $\Omega$ is given by  equation (\ref{newO}).
So we can  look for
solutions of the P$\Delta$E (\ref{hie2}) in the form of a combination of modulated waves (\ref{3.9}),
where the functions $\psi^{(s)}_{n,m}$ are  slowly varying functions on the lattice, 
i.e. $\psi^{(s)}_{n,m}=\psi^{(s)}_{n_1,m_1,m_2}$  and $\epsilon^{\gamma} = N^{-1}$.

Introducing the expansion (\ref{3.9}) in the Hietarinta equation (\ref{hie2}) and considering the
equations for $s=1$, $s=2$ and $s=0$ harmonics we deduce that the choice (\ref{3.10b}) is still valid. 
Moreover, the discrete slow varying variables $n_1$, $m_1$ and $m_2$ are defined in terms of $n$ and $m$ by equation (\ref{p31}).

Having fixed the constants $\beta_s$ we can now introduce the ansatz (\ref{3.9}) 
into equation (\ref{hie2}) and pick out the coefficients of the various harmonics $(E_{n,m})^s$ 
to get the determining equations. 

For $s=1$, having defined $\psi^{(1)}_{n,m} \doteq  \psi_{n,m}$, we obtain an
equation at the
first order in $\epsilon$ which is identically solved by the dispersion relation (\ref{real}). 

At $\epsilon^2$ we get the linear equation
\bea 
&& M_1 \, z \, [ o_1 (e_1-e_2)\Omega +e_1(o_1-e_2)] \,
(\psi_{n_1+1,m_1,m_2} - \psi_{n_1-1,m_1,m_2} )+  \nonumber \\ 
&+&M_2 \, \Omega \,[ o_1 (e_1-e_2)z - e_1(o_1-e_2)] \, (  \psi_{n_1,m_1+1,m_2} - \psi_{n_1,m_1-1,m_2} )
= 0, \nonumber
\eea
whose solution is given by 
\beq \nonumber
\psi_{n_1,m_1,m_2} = \phi_{n_2,m_2}, \qquad n_2 = n_1 - m_1.
\eeq
provided that the integers $M_1$ and $M_2$ are choosen as
\bea \label{m1m2h}
M_1 = S \, \Omega \,[ o_1 (e_1-e_2)z - e_1(o_1-e_2)], \qquad
M_2 = S \, z \, [ o_1 (e_1-e_2)\Omega +e_1(o_1-e_2)], 
\eea
where $S \in \mathbb{C}$ is a constant. Inserting $\Omega$ given by equation (\ref{newO}) in equation (\ref{m1m2h}) we can
show that the ratio $M_2/M_1$ coincides with the group velocity (\ref{gvh}).

At $\epsilon^3$ 
we get a nonlinear equation for $\phi_{n_2,m_2}$ which
depends on $\psi^{(0)}_{n_2,m_2}$ and $\psi^{(2)}_{n_2,m_2}$. It reads:
\bea \label{yy}
&& \phi_{n_2,m_2+1} - \phi_{n_2,m_2} +  
   c_1 \, ( \phi_{n_2+2,m_2} + \phi_{n_2-2,m_2} - 2 \, \phi_{n_2,m_2} )  + \nonumber  \\ 
&+& c_2 \,( \phi_{n_2+1,m_2} +
\phi_{n_2-1,m_2} - 2 \, \phi_{n_2,m_2} ) +
c_3  \, \phi_{n_2,m_2} |{\phi}_{n_2,m_2}|^2+  \nonumber  \\ 
&+& c_4  \, \psi^{(0)}_{n_2,m_2} \, {\phi}_{n_2,m_2}+
c_5  \, \psi^{(2)}_{n_2,m_2} \, \bar{{\phi}}_{n_2,m_2} = 0,
\eea
where the coefficients $c_i$, $1\leq i\leq 5$, depend on
$z$, $S$ and the lattice parameters $e_1,e_2,o_1$ and are given in Appendix B as  their expressions
are rather complicated.

The functions $\psi_{n_2,m_2}^{(0)}$ and $\psi_{n_2,m_2}^{(2)}$ that appear in equation (\ref{yy})
are obtained by considering the equations for the harmonics $s=0$, at the third order
in $\epsilon$, and $s=2$ 
at the second one. From them we get:
\beq \label{ni91h}
\psi_{n_2,m_2}^{(2)} = p_1 (\phi_{n_2,m_2})^2, 
\eeq
\beq \label{ni92h}
\psi_{n_2+1,m_2}^{(0)}  -  \psi_{n_2-1,m_2}^{(0)}= 
p_2\, [{\bar \phi}_{n_2,m_2} (\phi_{n_2+1,m_2} - \phi_{n_2-1,m_2}) + 
\phi_{n_2,m_2} ( {\bar  \phi}_{n_2+1,m_2} - {\bar \phi}_{n_2-1,m_2} ) ],
\eeq
with
$$
p_1=\frac{e_1 \,z -o_1}{e_1 \,o_1 (z-1)}, \qquad
p_2=\frac{e_1+o_1}{e_1 \, o_1 }.
$$
From equations (\ref{ni91h}) and (\ref{ni92h}) we evince that both $\psi_{n_2,m_2}^{(0)}$ and $\psi_{n_2,m_2}^{(2)}$ 
are expressed in term of $\phi_{n_2,m_2}$. In particular, we notice that $\psi_{n_2,m_2}^{(2)} $ depends from $\phi_{n_2,m_2}$ in a local way while
$\psi_{n_2,m_2}^{(0)} $ depends from $\phi_{n_2,m_2}$ in a non local way through a summation, namely
\beq \label{sol}
\psi_{n_2,m_2}^{(0)}=(-1)^n \left[ w_1 + 
p_2 \sum_{j=n_2}^\infty
(-1)^j({\bar \phi}_{j,m_2}\, \phi_{j+1,m_2}+ \phi_{j,m_2}\, {\bar \phi}_{j+1,m_2}) \right] +w_2,
\eeq
where $w_1,w_2$ are two arbitrary summation constants.

Inserting $\psi_{n_2,m_2}^{(2)}$ given by equation (\ref{ni91h}) in equation (\ref{yy}) we get
\bea \label{yyth}
&& \phi_{n_2,m_2+1} - \phi_{n_2,m_2} +  
   c_1 \, ( \phi_{n_2+2,m_2} + \phi_{n_2-2,m_2} - 2 \, \phi_{n_2,m_2} )  + \nonumber  \\ 
&+& c_2 \,( \phi_{n_2+1,m_2} +
\phi_{n_2-1,m_2} - 2 \, \phi_{n_2,m_2} ) +
\hat c_3  \, \phi_{n_2,m_2} |{\phi}_{n_2,m_2}|^2+ c_4  \, \psi^{(0)}_{n_2,m_2} \, {\phi}_{n_2,m_2}= 0,
\eea
where, using the form of the complex constant $S$, see Appendix A, and the fact that $z \doteq e^{{\rm {i}} \, k}$, the coefficients are:
\bea
&& c_1 = -\frac{P_2 [P_1 (\cos k +{\rm{i}}\sin k)+P_2]}
{4(P_2^2-P_1^2)}M_2^2 , \nonumber \\
&& c_2 = \frac{P_2 (P_1 \cos k +P_2)}
{P_2^2-P_1^2}M_2^2, \nonumber \\
&& \hat c_3 = \frac{2(P_1-P_2)[P_1(e_1-e_2)+P_2(e_2-o_1)](\cos k -1)}
{e_2 (o_1\,e_2+P_2)(P_1^2+P_2^2 + 2\, P_1 \, P_2 \, \cos k)}, \nonumber \\
&& c_4 = -\frac{2(P_1-P_2)(P_1-e_2^2+o_1\,e_2)(\cos k -1)}
{e_2 (P_1^2+P_2^2 + 2\, P_1 \, P_2 \, \cos k)}, \nonumber 
\eea
with
$$
P_1 = e_1(e_2-o_1), \qquad
P_2= o_1(e_1-e_2).
$$
Here $M_2$ is an arbitrary integer number, while $M_1$ is given by (see Appendix A)
$$
M_1=M_2 \frac{P_1^2+P_2^2 + 2\, P_1 \, P_2 \, \cos k}
{P_2^2-P_1^2}.
$$

%%%%%%%%%%%%%%%%%%%%%%%%%%%%%%%%%%%%%%%%%%%%%%%%%%%%%%%%%%%%%%%%%%%%%%%%%%%%%%%%%%%%%%%%%%%%%%%%%%%%%%%% 
%%%%%%%%%%%%%%%%%%%%%%%%%%%%%%%%%%%%%%%%%%%%%%%%%%%%%%%%%%%%%%%%%%%%%%%%%%%%%%%%%%%%%%%%%%%%%%%%%%%%%%%%
\subsection{Reduction of the lattice VKVM equation} \label{vkvm}
%%%%%%%%%%%%%%%%%%%%%%%%%%%%%%%%%%%%%%%%%%%%%%%%%%%%%%%%%%%%%%%%%%%%%%%%%%%%%%%%%%%%%%%%%%%%%%%%%%%%%%%% 
%%%%%%%%%%%%%%%%%%%%%%%%%%%%%%%%%%%%%%%%%%%%%%%%%%%%%%%%%%%%%%%%%%%%%%%%%%%%%%%%%%%%%%%%%%%%%%%%%%%%%%%%

The completely discrete version of the Volterra--Kac--Van Moerbeke (VKVM) equation is given by the following P$\Delta$E \cite{frank1}:
\beq \label{vkv}
\frac{u_{n,m+1}}{u_{n+1,m}}=
\frac{\alpha \, u_{n,m}-1}{\alpha \, u_{n+1,m+1}-1},
\eeq
Here $\alpha$ is a real lattice parameter and $u_{n,m}$ is a real field.

The dispersion relation of the linear part of equation (\ref{vkv}) is trivial. So,
we carry out the change of variable
$
u_{n,m} \mapsto
1 + u_{n,m}
$.
Then one can split equation (\ref{vkv}) into the linear and nonlinear parts:
\beq \label{vkv2}
\alpha (u_{n+1,m+1}-u_{n,m}) +(1-\alpha)(u_{n+1,m}-u_{n,m+1})=
\alpha (u_{n,m} \, u_{n+1,m}-u_{n+1,m+1}\,u_{n,m+1}).
\eeq

The dispersion relation for the linear waves is given by 
\beq \label{wwvkv}
\Omega = 
\frac{\alpha(z+1) -z}{\alpha(z+1) -1}, \qquad 
\omega = \arctan \left[
\frac{(2\alpha-1) \sin k} 
{(2\alpha^2-2\alpha+1)\cos k +2\alpha(\alpha-1)}
\right].
\eeq
From equations (\ref{wwvkv}),  
by differentiation with respect to $k$, we get the group velocity $\omega_{,k}$
\bea \label{gvhvkv}
\omega_{,k} =
\frac{(2\alpha-1)z}
{[\alpha(z+1) -z][\alpha(z+1) -1]} 
= \frac{(2\alpha-1)}
{2\alpha(\alpha-1)(\cos k+1) +1}.
\eea 

We now consider a solution of the P$\Delta$E (\ref{vkv2}) in the form of a combination of modulated waves, see
equation (\ref{3.9}), where $E_{n,m}$ is given by equation (\ref{3.8d}) with $\Omega$ as in equation (\ref{wwvkv}). As in the previous cases
the functions $\psi^{(s)}_{n,m}$ are to be slowly varying functions on the lattice, 
i.e. $\psi^{(s)}_{n,m}=\psi^{(s)}_{n_1,m_1,m_2}$  and $\epsilon^{\gamma} = N^{-1}$.

Introducing the expansion (\ref{3.9}) in the lattice VKVM equation (\ref{vkv2}) and considering the
equations for $s=1$, $s=2$ and $s=0$ we deduce that the choice (\ref{3.10b}) is still valid.
The discrete slow varying variables $n_1$, $m_1$ and $m_2$ are defined in terms of $n$ and $m$ by
the positions (\ref{p31}),
where $M_1,M_2 \in \mathbb{Z}$.

For $s=1$ we obtain an
equation at the
first order in $\epsilon$ which is identically solved by the dispersion relation (\ref{wwvkv}). 

At $\epsilon^2$ we get a linear equation
\bea 
&& M_1 \, z \,[\alpha(\Omega-1)+1] \,
(\psi_{n_1+1,m_1,m_2} - \psi_{n_1-1,m_1,m_2} )+  \nonumber \\ 
&+&M_2 \, \Omega \, [\alpha(z+1)-1] \, (  \psi_{n_1,m_1+1,m_2} - \psi_{n_1,m_1-1,m_2} )
= 0, \nonumber
\eea
whose solution is given by 
\beq \nonumber
\psi_{n_1,m_1,m_2} = \phi_{n_2,m_2}, \qquad n_2 = n_1 - m_1.
\eeq
provided that the integers $M_1$ and $M_2$ are choosen as
\bea \label{m1m2hvkvq}
M_1 = S \, \Omega[\alpha(z+1)-1], \qquad
M_2 = S \, z [\alpha(\Omega-1)+1], 
\eea
where $S \in \mathbb{C}$ is a constant. Inserting $\Omega$ given by equation (\ref{wwvkv}) 
in equation (\ref{m1m2hvkvq}) we get that the ratio $M_2/M_1$ coincides with the group velocity (\ref{gvhvkv}).
As shown in Appendix A it is possible to choose the complex constant $S$ 
in such a way that $M_1$ and $M_2$ are in fact integer numbers.

At $\epsilon^3$ 
we get a nonlinear equation for $\phi_{n_2,m_2}$ which
depends on $\psi^{(0)}_{n_2,m_2}$ and $\psi^{(2)}_{n_2,m_2}$. It reads:
\bea \label{yyvkv}
&& \phi_{n_2,m_2+1} - \phi_{n_2,m_2} +  
   c_1 \, ( \phi_{n_2+2,m_2} + \phi_{n_2-2,m_2} - 2 \, \phi_{n_2,m_2} )  + \nonumber  \\ 
&+& c_2 \,( \phi_{n_2+1,m_2} +
\phi_{n_2-1,m_2} - 2 \, \phi_{n_2,m_2} ) +
c_3  \, \psi^{(0)}_{n_2,m_2} \, {\phi}_{n_2,m_2}+
c_4  \, \psi^{(2)}_{n_2,m_2} \, \bar{{\phi}}_{n_2,m_2} = 0,
\eea
where the coefficients $c_i$, $1 \leq i \leq 4$, are:
\bea
&& c_1 = \alpha \, S^2 \, z^2 \, \frac{(1-2\alpha)[\alpha(z+1)-z]}{4[\alpha(z+1)-1]^2},\nonumber \\
&& c_2 = \alpha \, S^2 \, z \, \frac{(2\alpha-1)[\alpha(z+1)^2-z^2-1]}{2[\alpha(z+1)-1]^2},\nonumber \\
&& c_3 = \frac{\alpha(1-z^2)}{[\alpha(z+1)-z][\alpha(z+1)-1]},\nonumber \\
&& c_4 = \frac{\alpha(1-z^2)(z^2-z+1)}{[\alpha(z+1)-z][\alpha(z+1)-1]z}.\nonumber 
\eea

The functions $\psi_{n_2,m_2}^{(0)}$ and $\psi_{n_2,m_2}^{(2)}$ that appear in equation (\ref{yyvkv})
are obtained by considering the equations for the harmonics $s=0$, at the second order in
$\epsilon$, and $s=2$, at the third one.
 We get the following equations:
\beq \label{ni91hvkv}
\psi_{n_2,m_2}^{(2)} = p_1(\phi_{n_2,m_2})^2, 
\eeq
\beq \label{ni92hvkv}
\psi_{n_2+1,m_2}^{(0)}  -  \psi_{n_2-1,m_2}^{(0)}= 
p_2\, [{\bar \phi}_{n_2,m_2} (\phi_{n_2+1,m_2} - \phi_{n_2-1,m_2}) + 
\phi_{n_2,m_2} ( {\bar  \phi}_{n_2+1,m_2} - {\bar \phi}_{n_2-1,m_2} ) ],
\eeq
with
$$
p_1=\frac{(1-2\alpha)z}{\alpha(z+1)^2-z^2-1}, \qquad
p_2=\frac{2\alpha(1+z^2)-z^2-1}{(1-z^2)(\alpha-1)}.
$$
From equations (\ref{ni91hvkv}) and (\ref{ni92hvkv}) 
we evince that both $\psi_{n_2,m_2}^{(0)}$ and $\psi_{n_2,m_2}^{(2)}$ 
are expressed in term of $\phi_{n_2,m_2}$. In particular $\psi_{n_2,m_2}^{(0)}$ admits
a non local expansion as in equation (\ref{sol}).
Inserting $\psi_{n_2,m_2}^{(2)}$ given by equation (\ref{ni91hvkv}) in equation (\ref{yyvkv}) we get
\bea \label{yyt}
&& \phi_{n_2,m_2+1} - \phi_{n_2,m_2} +  
   c_1 \, ( \phi_{n_2+2,m_2} + \phi_{n_2-2,m_2} - 2 \, \phi_{n_2,m_2} )  + \nonumber  \\ 
&+& c_2 \,( \phi_{n_2+1,m_2} +
\phi_{n_2-1,m_2} - 2 \, \phi_{n_2,m_2} ) + c_3  \, \psi^{(0)}_{n_2,m_2} \, {\phi}_{n_2,m_2}+
\hat c_4  \, \phi_{n_2,m_2} |{\phi}_{n_2,m_2}|^2= 0,
\eea
where $\hat c_4=c_4 \, p_1$.

Let us write the coefficients that appear in equation (\ref{yyt}), i.e. $c_1,c_2,c_3, \hat c_4$,
using the form of the complex constant $S$, see Appendix A,  and the fact that $z \doteq e^{{\rm {i}} \, k}$. We get
\bea
&& c_1 = -\alpha \, \frac{(\alpha-1)(\cos k +{\rm{i}}\sin k)+\alpha}
{4(2\alpha-1)}M_2^2 , \nonumber \\
&& c_2 = \alpha \, \frac{(\alpha-1)\cos k+\alpha}
{2\alpha-1}M_2^2, \nonumber \\
&& c_3 = -{\rm{i}} \, \frac{2\alpha \sin k}
{2\alpha(\alpha-1)(\cos k+1) +1}, \nonumber \\
&& \hat c_4 = {\rm{i}} \, \frac{(2 \cos k-1)\sin k}
{(\alpha-1)(\cos k-1)[2\alpha(\alpha-1)(\cos k+1) +1]}. \nonumber 
\eea
As in the previous cases we can choose the integer number $M_2$, while 
$M_1$ is given by (see Appendix A)
$$
M_1=M_2 \frac{2\alpha(\alpha-1)(\cos k+1) +1}
{2\alpha-1}.
$$

%%%%%%%%%%%%%%%%%%%%%%%%%%%%%%%%%%%%%%%%%%%%%%%%%%%%%%%%%%%%%%%%%%%%%%%%%%%%%%%%%%%%%%%%%%%%%%%%%%%%%%%% 
%%%%%%%%%%%%%%%%%%%%%%%%%%%%%%%%%%%%%%%%%%%%%%%%%%%%%%%%%%%%%%%%%%%%%%%%%%%%%%%%%%%%%%%%%%%%%%%%%%%%%%%%
\subsection{Reduction of a non integrable lattice KdV equation} \label{nonintegrable}
%%%%%%%%%%%%%%%%%%%%%%%%%%%%%%%%%%%%%%%%%%%%%%%%%%%%%%%%%%%%%%%%%%%%%%%%%%%%%%%%%%%%%%%%%%%%%%%%%%%%%%%% 
%%%%%%%%%%%%%%%%%%%%%%%%%%%%%%%%%%%%%%%%%%%%%%%%%%%%%%%%%%%%%%%%%%%%%%%%%%%%%%%%%%%%%%%%%%%%%%%%%%%%%%%%

Let us now consider the following non integrable lattice KdV equation:
\beq \label{ni1}
u_{n,m+1} - u_{n,m-1} = 
\frac{\alpha}{4} ( u_{n+3,m}-3u_{n+1,m}+3 u_{n-1,m}-u_{n-3,m}) + \beta [ (u_{n+1,m})^2 - (u_{n-1,m})^2],
\eeq
where $\alpha, \beta \in \mathbb{R}$ are the lattice parameters and $u_{n,m}$ is a real field. 

As we did for previous cases  we
apply the standard discrete Fourier transform procedure introducing
$u_{n,m} = z^n \,  \Omega^m$ into  the linear part of the lattice equation (\ref{ni1}). 
Here  
$z \doteq e^{{\rm {i}} \, k}$ and $\Omega \doteq  e^{-{\rm {i}} \,  \omega}$.
We easily get:
\beq \label{ni3}
\Omega - \Omega^{-1} = \frac{\alpha}{4} ( z - z^{-1})^3.
\eeq
Hence the dispersion relation reads
\beq \nonumber
\omega = \arcsin \left( \alpha \sin^3 k \right)
\eeq
and the corresponding group velocity is
\beq \label{ni5dd}
\omega_{,k} = - \frac{3}{4}  \frac{\alpha \, \Omega}{1 + \Omega^2}
\frac{(z^4 - 1)(z^2 - 1)}{z^3}=
\frac{3 \,\alpha  \cos k \sin^2 k}{\sqrt{1-\alpha^2 \sin^6 k}}. 
\eeq

Introducing the expansion (\ref{3.9}) into the P$\Delta$E (\ref{ni1}), 
where $E_{n,m}$ is given by equation (\ref{3.8d}) and taking into account that
\beq \nonumber
f_{n\pm k} = g_{n_1} \pm 
\frac{k}{2 N} ( g_{n_1 +1} - g_{n_1 - 1} ) + 
\frac{k^2}{4 N^2} ( g_{n_1 +1} + g_{n_1 - 1} - 2 g_{n_1}) + 
O(N^{-3}),
\eeq
we get the standard choice (\ref{3.10b}).

Let us  consider now the equations for the harmonics $s=1$. 
The equation at the order $\epsilon$ is identically satisfied by taking into account the dispersion relation 
(\ref{ni3}). The equation at the order $\epsilon^2$ is satisfied if we introduce the index $n_2 = n_1 - m_1$ when
$M_1$ and $M_2$ are choosen as 
\beq
M_1 = S  \left(\Omega+\frac{1}{\Omega}\right), \qquad
M_2 = - \frac{3}{4} \, S \, \alpha \, \frac{(z^4 - 1)(z^2 - 1)}{z^3}. \label{nuuu}
\eeq
We notice that the group velocity $\omega_{,k}$ (\ref{ni5dd}) 
coincides again with the ratio $M_2/M_1$, as in equation (\ref{l3}).
From equation (\ref{nuuu}), 
using the fact that $z \doteq e^{{\rm {i}} \, k}$ and $\Omega \doteq e^{-{\rm {i}} \, \omega}$ , we obtain
$$
M_1= -2 S \cos \omega, \qquad
M_2= -6 S \alpha \cos k \sin^2 k.
$$
We can now fix the (real) constant $S$ in a such a way that $M_1$ is an integer number;
$M_2$ will be an integer if the group velocity is a rational number. Hence not all
values of $k$ are admissible, but only those which make $\omega_{,k}$ (\ref{ni5dd}) rational.

The equation at the order $\epsilon^3$ is given by
\beq \label{ni8}
\phi_{n_2, m_2+1} - \phi_{n_2, m_2} + 
c_1 \,( \phi_{n_2+1,m_2} +  \phi_{n_2-1,m_2} - 2  \phi_{n_2,m_2} ) + c_2 (\, \phi_{n_2,m_2}\,   \phi_{n_2,m_2}^{(0)} + 
{\bar \phi}_{n_2,m_2} \,  \phi_{n_2,m_2}^{(2)} )= 0,  
\eeq
where $\phi_{n_2,m_2}=\psi_{n_1,m_1,m_2}, n_2 = n_1 - m_1$ and 
$c_1,c_2$ are known, easy to compute but too complicate to write down,  complex coefficients depending on $z$
and on the lattice parameter $\alpha$. 
The functions $\psi_{n_2,m_2}^{(0)}$ and $\psi_{n_2,m_2}^{(2)}$ that appear in equation (\ref{ni8})
are obtained by considering the equations for the harmonics $s=0$, at the third order
in $\epsilon$, and $s=2$, 
at the second one. We get
the following equations:
\beq \label{ni91}
\psi_{n_2,m_2}^{(2)} = p_1 \, (\phi_{n_2,m_2})^2, 
\eeq
\beq \label{ni92}
\psi_{n_2+1,m_2}^{(0)}  -  \psi_{n_2-1,m_2}^{(0)}= 
p_2  \, [{\bar \phi}_{n_2,m_2} (\phi_{n_2+1,m_2} - \phi_{n_2-1,m_2}) + 
\phi_{n_2,m_2} ( {\bar  \phi}_{n_2+1,m_2} - {\bar \phi}_{n_2-1,m_2} ) ],
\eeq
where $p_1,p_2$ are known complex coefficients depending on $z$ and on the lattice parameters.
From equations (\ref{ni91}) and (\ref{ni92}) 
we evince that both $\psi_{n_2,m_2}^{(0)}$ and $\psi_{n_2,m_2}^{(2)}$ 
are expressed in term of $\phi_{n_2,m_2}$. As in the previous cases $\psi_{n_2,m_2}^{(0)}$
admits a non local expansion as in equation (\ref{sol}).
Hence the P$\Delta$E (\ref{ni8}) is a well defined
lattice equation in the field variable $\phi_{n_2,m_2}$.

%%%%%%%%%%%%%%%%%%%%%%%%%%%%%%%%%%%%%%%%%%%%%%%%%%%%%%%%%%%%%%%%%%%%%%%%%%%%%%%%%%%%%%%%%%%%%%%%%%%%%%%% 
%%%%%%%%%%%%%%%%%%%%%%%%%%%%%%%%%%%%%%%%%%%%%%%%%%%%%%%%%%%%%%%%%%%%%%%%%%%%%%%%%%%%%%%%%%%%%%%%%%%%%%%%
\section{Conclusive remarks} \label{conclusions}
%%%%%%%%%%%%%%%%%%%%%%%%%%%%%%%%%%%%%%%%%%%%%%%%%%%%%%%%%%%%%%%%%%%%%%%%%%%%%%%%%%%%%%%%%%%%%%%%%%%%%%%% 
%%%%%%%%%%%%%%%%%%%%%%%%%%%%%%%%%%%%%%%%%%%%%%%%%%%%%%%%%%%%%%%%%%%%%%%%%%%%%%%%%%%%%%%%%%%%%%%%%%%%%%%%
In this paper we have shown that we can construct a well defined procedure to carry out the reductive perturbation technique on the lattice. In this case, at difference with respect to the differential--difference case, we are able to solve all linear equations and thus can obtain a final nonlinear difference equation. To do so we had to apply some non trivial but at the end obvious tricks which consist in the introduction 
of appropriate lattice variables so as to be able to perform the symmetric reduction of the linear
discrete wave equation.  

Applying the perturbative--reductive technique  to some integrable and non integrable equations we obtain some new completely discrete NLS equations. As some of these equations (\ref{yyy}), (\ref{yyth}) and (\ref{yyt}) come from the reduction of integrable equations we expect them to be also integrable. However they are very different from the Ablowitz--Ladik discrete--discrete  NLS
\cite{al} as all contains, apart from the nearest neighboring points, also the points $n \pm 2$ and either they are completely local or they have non local completely irregular terms (depending on $(-1)^n$). 

So we are at the moment, from one side  extending our analysis to other well known integrable equations, like the discrete time Toda lattice, the sine--Gordon and the Volterra equations and from the other
using the integrability properties of the starting nonlinear equations (i.e. Lax pairs or generalized symmetries) to show the integrability of the derived equations. If our equation are integrable than we have presented a very important tool for obtaining new integrable equations and for analyzing the far field behavior of physical problems described by differential--difference or partial difference equations.

In the derivation we introduced the request that the far field expansion of a slow varying function on the lattice should depend on the discrete asymptotic variables in a symmetric way. As a consequence of this ansatz we got that the non local resulting equation depends on $(-1)^n$. This may not be a necessary ansatz
and work is in progress in this direction.

%%%%%%%%%%%%%%%%%%%%%%%%%%%%%%%%%%%%%%%%%%%%%%%%%%%%%%%%%%%%%%%%%%%%%%%%%%%%%%%%%%%%%%%%%%%%%%%%%%%%%%%% 
%%%%%%%%%%%%%%%%%%%%%%%%%%%%%%%%%%%%%%%%%%%%%%%%%%%%%%%%%%%%%%%%%%%%%%%%%%%%%%%%%%%%%%%%%%%%%%%%%%%%%%%%
\section*{Acknowledgments}
%%%%%%%%%%%%%%%%%%%%%%%%%%%%%%%%%%%%%%%%%%%%%%%%%%%%%%%%%%%%%%%%%%%%%%%%%%%%%%%%%%%%%%%%%%%%%%%%%%%%%%%% 
%%%%%%%%%%%%%%%%%%%%%%%%%%%%%%%%%%%%%%%%%%%%%%%%%%%%%%%%%%%%%%%%%%%%%%%%%%%%%%%%%%%%%%%%%%%%%%%%%%%%%%%%

The author D.L. wishes to thank  X-D. Ji for interesting and fruitful discussions. The author M.P. 
wishes to express his gratitude to 
G. Satta for his many suggestions during the writing of part of this manuscript.
MP wishes also to aknowledge F. Musso and O. Ragnisco for helpful
discussions.  D.L.  was partially supported by  PRIN Project ``SINTESI-2004'' of the  Italian Minister for  Education and Scientific Research and from  the Projects {\sl Sistemi dinamici nonlineari discreti:
simmetrie ed integrabilit\'a} and {\it Simmetria e riduzione di equazioni differenziali di interesse fisico-matematico} of GNFM--INdAM.

%%%%%%%%%%%%%%%%%%%%%%%%%%%%%%%%%%%%%%%%%%%%%%%%%%%%%%%%%%%%%%%%%%%%%%%%%%%%%%%%%%%%%%%%%%%%%%%%%%%%%%%% 
%%%%%%%%%%%%%%%%%%%%%%%%%%%%%%%%%%%%%%%%%%%%%%%%%%%%%%%%%%%%%%%%%%%%%%%%%%%%%%%%%%%%%%%%%%%%%%%%%%%%%%%%
\section*{Appendix A} 
%%%%%%%%%%%%%%%%%%%%%%%%%%%%%%%%%%%%%%%%%%%%%%%%%%%%%%%%%%%%%%%%%%%%%%%%%%%%%%%%%%%%%%%%%%%%%%%%%%%%%%%% 
%%%%%%%%%%%%%%%%%%%%%%%%%%%%%%%%%%%%%%%%%%%%%%%%%%%%%%%%%%%%%%%%%%%%%%%%%%%%%%%%%%%%%%%%%%%%%%%%%%%%%%%%

From equations (\ref{m1m2}), (\ref{m1m2h}) and (\ref{m1m2hvkvq}) we get that the
coefficients $M_1$ and $M_2$ can always be written in the following general form:
\beq
M_1 =S \, \Omega ( P z - Q), \qquad M_2 =S \,z  ( P \,\Omega + Q), \label{Mgen}
\eeq
where $S \in \mathbb{C}$ is a suitable constant such that $M_1,M_2 \in \mathbb{Z}$,
$\Omega \doteq e^{-{\rm {i}}\omega}$, $z \doteq e^{{\rm {i}}k}$ and $P,Q \in \mathbb{R}$ are given by:
\bea
P= p-q, \quad Q= p+q \qquad && {\rm {lattice~mKdV~equation}}, \nonumber \\
P=o_1(e_1-e_2), \quad Q=e_1(o_1-e_2) 
\qquad &&{\rm {Hietarinta~equation}}, \nonumber \\
P= \alpha, \quad Q=1-\alpha \qquad &&{\rm {lattice~VKVM~equation}}. \nonumber
\eea
The (real) dispersion relation for the linear parts of the above lattice equations can be
written in the form
\beq
\Omega=\frac{P-Q\,z}{P\,z-Q}. \label{PQ}
\eeq
Let us now define the complex constant $S$ 
as $S \doteq \rho \,  e^{{\rm {i}} \, \theta}$, with $\rho \in \mathbb{R}_+$ and
$-\pi \leq \theta < \pi$. 
From equations (\ref{Mgen}) and (\ref{PQ}) we get
\bea 
&& {\rm {Re}}(M_1) = \rho \,[ P \cos (\theta) -Q \cos (\theta+k)], \label{I} \\
&& {\rm {Im}}(M_1) = \rho \,[ P \sin (\theta) -Q \sin (\theta+k)], \label{II}  \\
&& {\rm {Re}}(M_2) = \frac{\rho  \,(P^2-Q^2)}{P^2+Q^2-2 P Q \cos k}
[ P \cos (\theta) -Q \cos (\theta+k)], \label{III}  \\
&& {\rm {Im}}(M_2) = \frac{\rho \, (P^2-Q^2)}{P^2+Q^2-2 P Q \cos k}
[ P \sin (\theta) -Q \sin (\theta+k)]. \label{IIII}  
\eea
Since $M_1, M_2 \in \mathbb{Z}$ 
we have to require that ${\rm {Im}}(M_1)={\rm {Im}}(M_2)=0$.
From equations (\ref{II}) and (\ref{IIII}) we obtain 
\beq \label{conI}
\theta = -\arctan \left(   
\frac{Q \sin k}{Q\cos k -P}
\right)+ \ell\, \pi, \qquad \ell \in \mathbb{Z}.
\eeq
We have now to require that $M_1={\rm {Re}}(M_1),M_2={\rm {Re}}(M_2)  \in \mathbb{Z}$. 
According to equations (\ref{I}), (\ref{III}) and (\ref{conI}) we get
\bea
&& M_1= (-1)^\ell \rho \, (P^2+Q^2-2 P Q \cos k)^{1/2}, \label{re1} \\
&& M_2 = (-1)^\ell \rho \,
\frac{(P^2-Q^2)}{(P^2+Q^2-2 P Q \cos k)^{1/2}}.\label{re2}
\eea
We can fix arbitrarily the integer number $M_2$,
(or equivalently $M_1$) and express
$M_1$ (or $M_2$) in terms of it:
\beq \label{uu}
M_1=M_2\frac{P^2+Q^2-2 P Q \cos k}{P^2-Q^2}.
\eeq
From (\ref{uu}) we can see that not all values of $k$ are admissible since 
$M_1$ has to be integer.

Let us finally notice that the fact that equation (\ref{uu}) contains $\cos(k)$ implies 
that the ratios $M_2/M_1$ and $P/Q$ 
are constrained as $-1\, \leq \, \cos(k) \, \leq 1$.
We have the following cases (see Figure 1):
\bea
P/Q \in (1,\infty) &\quad \Rightarrow \quad &
M_2/M_1 \in [(P/Q-1)/(P/Q+1),(P/Q+1)/(P/Q-1)] 
\in \mathbb{Q}, \nonumber \\ 
P/Q \in (0,1) &\quad \Rightarrow \quad &
M_2/M_1\in [(P/Q+1)/(P/Q-1),(P/Q-1)/(P/Q+1)] 
\in \mathbb{Q}, \nonumber \\ 
P/Q\in (-1,0) &\quad \Rightarrow \quad &
M_2/M_1\in [(P/Q-1)/(P/Q+1),(P/Q+1)/(P/Q-1)] 
\in \mathbb{Q}, \nonumber \\ 
P/Q\in (-\infty,-1) &\quad \Rightarrow \quad &
M_2/M_1\in [(P/Q+1)/(P/Q-1),(P/Q-1)/(P/Q+1)] 
\in \mathbb{Q}. \nonumber  
\eea

\begin{figure}[h!]

\begin{center}
\includegraphics[height=10cm]{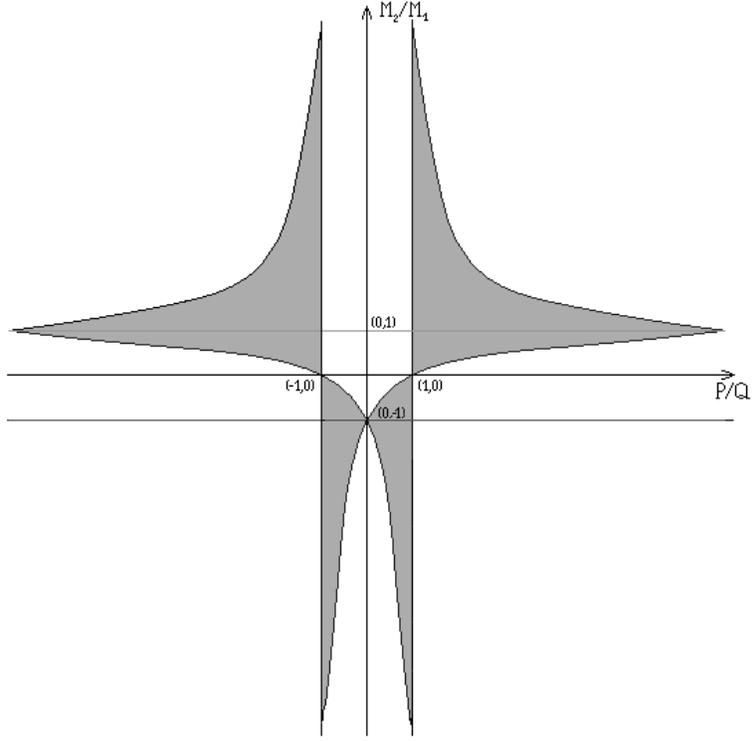}
\end{center}
\caption{The grey zones denote the allowed regions for the ratio $M_2/M_1$ in terms of the ratio $P/Q$}
\end{figure}

%%%%%%%%%%%%%%%%%%%%%%%%%%%%%%%%%%%%%%%%%%%%%%%%%%%%%%%%%%%%%%%%%%%%%%%%%%%%%%%%%%%%%%%%%%%%%%%%%%%%%%%% 
%%%%%%%%%%%%%%%%%%%%%%%%%%%%%%%%%%%%%%%%%%%%%%%%%%%%%%%%%%%%%%%%%%%%%%%%%%%%%%%%%%%%%%%%%%%%%%%%%%%%%%%%
\section*{Appendix B} 
%%%%%%%%%%%%%%%%%%%%%%%%%%%%%%%%%%%%%%%%%%%%%%%%%%%%%%%%%%%%%%%%%%%%%%%%%%%%%%%%%%%%%%%%%%%%%%%%%%%%%%%% 
%%%%%%%%%%%%%%%%%%%%%%%%%%%%%%%%%%%%%%%%%%%%%%%%%%%%%%%%%%%%%%%%%%%%%%%%%%%%%%%%%%%%%%%%%%%%%%%%%%%%%%%%

The coefficients $c_i$, $1 \leq i \leq 5$ that appear in equation (\ref{yy}) are
\bea
&& c_1  = S^2 \, z^2 \, 
\frac{P_2(P_1^2-P_2^2)(P_1 z +P_2)}
{4( P_1+P_2 z )^2}, \nonumber \\
&& c_2 = -S^2 \, z \, 
\frac{P_2(P_1^2-P_2^2)[P_1(1+z^2) +2\,P_2 z]}
{2( P_1+P_2 z )^2}, \nonumber \\
&&c_3 = \frac{(z-1)(P_1-P_2)
[Q_1 z^5+Q_2 z^4+ Q_3 z^3 +Q_4 z^2 +Q_5 z +Q_6]}
{e_2 (P_1 -e_1 \,e_2)( P_1 +P_2 z)^2( P_1 z +P_2)^2 z}, \nonumber \\
&& c_4 = \frac{(z-1)^2(P_1-P_2)(e_2^2-e_1\,e_2 +P_2)}
{e_2( P_2 z +P_1)( P_1 z +P_2)}, \nonumber \\
&& c_5 = \frac{(z-1)^2(P_1-P_2)
[R_1 z^4+R_2 z^3+R_3 z +R_4]}
{e_2 ( P_1 +P_2 z)^2( P_1 z +P_2)^2 z}, \nonumber
\eea
with
\bea
&& \left\{
\begin{array}{l}
P_1 = e_1(e_2-o_1), \\
P_2= o_1(e_1-e_2), 
\end{array} \right. \nonumber \\ 
&& \left\{
\begin{array}{l}
Q_1 = P_1 \,P_2 (P_1\,e_1 +P_2\,e_2),  \\
Q_2 = P_1^3(e_1-e_2)+P_2^3(e_2-o_1)+P_1\,P_2 (P_2\,e_1 +2P_1 \,e_2-P_1\,o_1),   \\
Q_3 = - P_1[P_1^2(e_1-e_2)+P_2^2(e_1+4o_1 -3\,e_2)+P_1\,P_2(3e_2 -e_1)],  \\
Q_4 = - P_2[P_1^2(4e_1-3e_2+o_1)+P_2^2(o_1 -e_2)+P_1\,P_2(3e_2 -o_1)],  \\
Q_5 = -P_1^3(e_1-e_2)-P_2^3(e_2-o_1)-P_1\,P_2 (P_2\,e_1 -2P_1\,e_2 -P_1\,o_1),  \\
Q_6 = P_1 \,P_2 (P_1\,e_2 +P_2\,o_1),
\end{array} \right. \nonumber \\  
&& \left\{
\begin{array}{l}
R_1 = P_2[P_1^2+P_2^2+P_1\,P_2 +P_2(e_2^2-e_1\,e_2)], \\
R_2 = P_2^3+(e_2^2-e_1\,e_2)(P_2^2-P_1^2)+P_1\,P_2 (e_2^2-e_1\,e_2 +P_1 +3P_2), \\
R_3 = -P_2^3- (e_2^2-e_1\,e_2)(P_2^2-P_1^2)- P_1\,P_2 (e_2^2-e_1\,e_2 +P_1 +P_2),  \\
R_4 = P_1(P_1 \, e_2^2 -P_2^2 -P_1 \,e_1\,e_2) . 
\end{array} \right. \nonumber
\eea
%The coefficient $\hat c_3$ that appears in equation (\ref{yyt}) reads
%$$
%\hat c_3 = \frac{(z-1)^2 (P_1-P_2)[P_1(e_1-e_2)+P_2(e_2-o_1)]}
%{e_2( P_2 z +P_1)( P_1 z +P_2)(e_1\,e_2-P_1)}.
%$$

%%%%%%%%%%%%%%%%%%%%%%%%%%%%%%%%%%%%%%%%%%%%%%%%%%%%%%%%%%%%%%%%%%%%%%%%%%%%%%%%%%%%%%%%%%%%%%%%%%%%%%%% 
%%%%%%%%%%%%%%%%%%%%%%%%%%%%%%%%%%%%%%%%%%%%%%%%%%%%%%%%%%%%%%%%%%%%%%%%%%%%%%%%%%%%%%%%%%%%%%%%%%%%%%%%

\end{document}